\documentclass[referee]{aa} 
\usepackage{graphicx}
\begin{document}
   \title{Pseudo-Newtonian potential for the relativistic accretion disk around
rotating central objects having hard surface}


   \author{Shubhrangshu Ghosh
        } 
          

   \offprints{Shubhrangshu Ghosh}

   \institute{Department of Physics and Centre for Theoretical Studies,
                Indian Institute of Technology, Kharagpur-721302, India \\
               \email{subharang@phy.iitkgp.ernet.in}
             }

   \date{Received ; accepted }

\abstract{ 
Here we furnish a pseudo-Newtonian potential for accretion disk modeling around
rotating central objects having hard surface. That means, the potential may be useful
to describe the accretion disk around rotating neutron stars and strange stars. The potential
can describe the general relativistic effects in accretion disk properly, essentially
studying the relativistic hydrodynamics of disk. As our intention lies in
understanding the fluid dynamics of the accretion disk around rotating 
central objects with hard surface, it is necessary to incorporate the rotation of the central object
which affects the inner edge of the disk. We derive our potential, starting
from the Hartle Thorne metric. The potential can reproduce all the
essential properties of general relativity with $10\%$ error at most,
for slowly rotating neutron/strange stars.}
 
\maketitle
\keywords {accretion disk -- pseudo-potential --
                rotating compact object -- gravitation -- relativity
               }

\markboth{Pseudo-potential for the accretion disk around rotating hard surface
}{ S. Ghosh }

\maketitle

%

\section{Introduction}

It is an established fact in astronomy and astrophysics that  
one of the method to study the accretion disk phenomenon is accomplished 
through the approach of pseudo-Newtonian method which 
the physicists prefer to avoid the complex and cumbersome general relativistic equations due to their
effectiveness in the mimicry of the geometry of space-time. 
This model was first initiated by Shakura \& Sunyaev (1973) when he proposed the simple 
Newtonian potential for non-rotating black holes. As the
relativistic effects are extremely important near the black holes, this 
potential lacks the luster of describing the essential inner
properties of the disk.

After that, Paczy\'nski \& Wiita (1980) modified this Newtonian potential in 
conformity with the Schwarzschild geometry, which can naturally reveal approximately all 
properties of the disk around non-rotating black holes, even that of the inner
most part of the disk. The potential can reproduce the marginally stable orbit $(r_s)$,
marginally bound orbit $(r_b)$ and efficiency per unit mass at the
last stable circular orbit $(E_s)$. The last parameter agrees 
with most a $10\%$ error but the first two cases exactly match with that of the metric.
Not only that the relativistic fluid dynamics of the accretion disk can be studied using this
pseudo-potential, the other properties of the disk like the spectral studies
can also be fulfilled around non-rotating black holes.  
After a gap of ten years, Nowak \& Wagoner (1991) proposed  
another potential for accretion disk around non-rotating black holes. This potential
also can mimic most of the properties of the disk governed by Schwarzschild geometry. However,
epicyclic frequency of the disk can be best analyzed with the potential in concern.

Artemova et al. (1996) proposed a couple of pseudo-potentials, which can describe 
the accretion disk around rotating black holes. These potentials are well analyzed by a
number of astrophysicists later. Very recently, Mukhopadhyay (2002a) and
Mukhopadhyay \& Misra (2003) prescribed some new potentials,
using which properties of accretion disk can be very well described in 
Kerr geometry. The more interesting fact, however, lies in the methodology adopted by 
Mukhopadhyay (2002a) in describing the accretion disk around 
rotating black holes which can be used to derive the pseudo-potential for any metric
according to the physics concern. The potentials proposed by Mukhopadhyay \& Misra (2003)
can be used for the time-dependent simulation of accretion disk around rotating black holes
and neutron stars. Except the potential described by Mukhopadhyay (2002a), most of the potentials
were proposed by hand. 

All the above mentioned potentials are meant for
accretion disk around black holes. However, herein, our course of study would lay in the
derivation of a pseudo-potential meant for accretion disk around rotating
central objects with a hard surface, which mainly consist of neutron stars, strange stars,
white dwarfs and other highly gravitating stars. As the radius
of White dwarf is very large, the general relativity is not of much
importance in the accretion disk around it.
The basic differences between the accretion disk having centrally
hard surface and that of horizon are (Mukhopadhyay 2002b) its
(1) inner boundary condition, (2) metric, (3) formation of shock
and (4) luminosity of the disk. In case of hard surface, inflowing
matter has to be stopped at the extreme inner edge of the disk as
there is an extended object whose space-time metric is 
described by Hartle \& Thorne (1968). However for non-rotating case the metric 
reduces to Schwarzschild geometry. Hence, it is quite obvious that the 
geometry of space time remains the same for central objects having hard
surface and event horizon, when they are non-rotating. Hereinafter, we will call the
class of central objects having hard surface as GOHS to distinguish it from black hole.
Around the black hole, only one stable shock formation is possible. In the case of
accretion disk around GOHS, the formation of the shock in a particular matter flow (upto stellar surface) 
may happen twice in certain situations (Mukhopadhyay 2002b). As far as the luminosity
is concerned, it is greater in case of accretion disk around
GOHS than that of a black hole, which mostly describes the 
difference in the nucleosynthesis between them.

Though various pseudo-potentials are known for describing the various
accretion disk properties of the central objects with event horizon
(rotating or non-rotating black holes), which can approximately
mimic the general-relativistic space-times, till date no such potential that can
describe relativistic fluid dynamical properties of the accretion disk around
rotating GOHS is known to the author. Prasanna \& Mukhopadhyay (2003) 
studied the general set of fluid equations with the inclusion of the effect
of Coriolis force on accretion flows around rotating compact objects. Still works
have been carried out in this field, sometimes using Paczy\'nski \& 
Wiita (1980) potential and  often, by numerical simulation. Hence, 
it intended the author to study the relativistic hydrodynamics of the accretion 
disk around GOHS, with the inclusion of a proper pseudo-Newtonian-potential.

It was known from various observational works that cosmic objects in 
our nature are spinning. Still we must convince ourselves by the analysis of  
specific phenomena, which proved that rotation of the central objects 
should be taken into account to study the properties of the inner edge of 
its accretion disk. Iwasawa et al. (1996) showed in the context of 
their observation of variable iron K emission line in MCG-6-30-15 that, it is arising from the 
inner part of the accretion disk and is strongly related with the rotation of the black hole. 
As the inner region of the accretion disk is very much influenced by the rotation of 
the central object, it should be incorporated into the theoretical studies. 
From another observational point of view (Karas \& Kraus 1996; Iwasawa et al. 1996), 
it was found that central black holes in galactic nuclei are rapidly rotating. 
Pulsars are well observed in our Universe which are nothing but rapidly 
rotating neutron stars. One more observational fact was established 
that the tidal excitations in coalescing binaries involves rapidly
exciting neutron star. Gravity wave oscillations from neutron stars 
enstrength the fact that neutron stars are rapidly spinning (Fragile, Mathews \& 
Wilson 2001). The occurrence of the nearly coherent brightness oscillations during 
thermonuclear X-ray bursts from neutron star low-mass X-ray binaries with the 
rotation period of these neutron stars, enhance the possibility that 
these are spinning fast enough to become unstable to gravity wave radiation.

All these and more, established the fact that compact objects and other cosmic
objects in our nature are rotating. As it has been mentioned above that no proper pseudo-potential
has yet been established to describe the accretion
flows around rotating neutron stars or in general for any GOHS, we make deliberate attempt to
describe a pseudo-potential that can mimic the geometry of space-time of  
rotating hard surfaces and can be further used to study the global solutions
of their accretion disk. The author found it more convincing in the work 
of Mukhopadhyay (2002a) in which, the pseudo-potential was derived for the 
rotating black holes starting from the Kerr metric. Most of the other 
potentials for the rotating black holes are proposed by hand, there are no 
connection with the space-time metric. Here, we follow the same method as 
Mukhopadhyay (2002a) to derive the potential for the accretion disk around 
GOHS, starting from the Hartle-Thorne (HT) metric
(Hartle \& Thorne 1968). As the HT metric is restricted for slow rotations only, 
our result will be valid only for the cases of slow rotation. As the metric
is involved directly to our calculation, it has been easy to reproduce most
of the features of Hartle-Thorne geometry by our potential. The potential
reduces to that of the Paczy\'nski-Wiita for non rotating case. 

As these methods 
(pseudo-Newtonian potential) are approximate in nature, it can be so that values 
of $r_b$, $r_s$ and $E_s$ for both HT and Kerr geometry may be quite 
near to each other at different values of $j$, which may  motivate someone 
to use the pseudo-potential described by Mukhopadhyay (2002a), meant 
for Kerr geometry, to describe approximately the accretion disk around 
rotating hard surface. But it should be kept in mind that Kerr geometry 
can not describe the solution in the interior as it is meant for event 
horizon. On the other hand HT geometry can describe it. As, we are 
considering here the rotating hard surface (and not event horizon), for the 
continuity of the solution from disk to inside the star, pseudo-Newtonian potential that can mimic HT metric 
has only to be taken, since the potential corresponding to Kerr metric Mukhopadhyay (2002a) is valid 
only up to even horizon. In the next section we elucidate 
the derivation of our said potential. In \S 3, we will compare a few results of the HT geometry with that
of our potential. In \S 4, we will make our conclusions.

\section{Basic equations and pseudo-potential}

The Lagrangian density for a particle in the Hartle Thorne space-time (neglecting
the higher quadrapole terms) (Hartle \& Thorne 1968), at the equatorial 
plane ($\theta=\pi/2$) can be written as
\begin{equation}
2{\cal L}=-\left(1-\frac{2M}{r}-\frac{2j^2}{r^4}\right){\dot 
t}^2+\left(1-\frac{2M}{r}+\frac{2j^2}{r^4}\right)^{-1}{\dot r}^2
+r^2{\dot \phi}^2-\frac{4j}{r}{\dot \phi}{\dot t}
\label{1}
\end{equation}
where, over-dots denote the derivative with respect to the proper-time $\tau$ and
$j$ denotes the angular momentum of the central object. As the metric is valid only
for slowly rotating stars, the rotation parameter $j$ is restricted, may be at 
say, $j\leq 0.5$. 
The geodesic equations of motion are 
\begin{equation}
E={\rm constant}=\left(1-\frac{2M}{r}-\frac{2j^2}{r^4}\right){\dot 
t}+\frac{2j}{r}{\dot \phi},
\label{2}
\end{equation} 

\begin{equation}
L={\rm constant}=r^2{\dot \phi}-\frac{2j}{r}{\dot t}.
\label{3}
\end{equation}
For the particle with non-zero rest mass, $g_{\mu\nu}p^\mu p^\nu=-m^2$ 
(where, $p^\mu$ is the momentum of the particles and $g_{\mu\nu}$ is the metric).   
Replacing the solution for $\dot t$ and $\dot \phi$ from (\ref{2}) and (\ref{3}) into 
(\ref{1}), we get a differential equation for $r$,   
\begin{equation}
\left(\frac{dr}{d\tau}\right)^2=\left(\frac{4hj^2}{gr^6}
+\frac{16j^4}{gr^{10}}-\frac{g}{r^2}\right)L^2+\left(\frac{h}{g}
+\frac{4j^2}{gr^4}\right)E^2+\left(-\frac{4hj}{gr^3}
-\frac{16j^3}{gr^7}\right)EL-gm^2=\Psi,
\label{4}
\end{equation}
where, $g=\left(1-\frac{2m}{r}+\frac{2j^2}{r^4}\right)$, $h=
\left(1-\frac{2m}{r}-\frac{2j^2}{r^4}\right)$. Here $\Psi$ can be identified as an 
effective potential for the radial geodesic motion.
The conditions for circular orbits are
\begin{equation}
\Psi=0,\hskip1.cm\frac{d\Psi}{dr}=0.
\label{5}
\end{equation}
Solving for $E$ and $L$ from (\ref{5}), we get
\begin{eqnarray}
\nonumber
E&=&\frac{1}{m^2r^3(-4j^2+Mr^3)}[(2j^3m^2+jm^2(4M-3r)r^3
-\sqrt{m^4(5j^2+Mr^3)(2j^2+r^3(-2M+r))^2})\\
\nonumber
&&\{(-12j^4m^2r^2+2j^2m^2(9M-7r)r^5+m^2M(3M-r)r^8+
6jr^2\sqrt{m^4(5j^2+Mr^3)(2j^2+r^3(-2M+r))^2})/\\
&&(36j^4-r^6(-3M+r)^2+12j^2r^3(-3M+2r))\}^{1/2}],
\label{6}
\end{eqnarray}

\begin{eqnarray}
\nonumber
L&=&\frac{1}{m^2r^3(-4j^2+Mr^3)}[(2j^3m^2+jm^2(4M-3r)r^3
-\sqrt{m^4(5j^2+Mr^3)(2j^2+r^3(-2M+r))^2})\\
\nonumber
&&\{(-12j^4m^2r^2+2j^2m^2(9M-7r)r^5+m^2M(3M-r)r^8+6jr^2\sqrt{m^4(5j^2+Mr^3)(2j^2+r^3(-2M+r))^2})/\\
\nonumber
&&(36j^4-r^6(-3M+r)^2+12j^2r^3(-3M+2r))\}^{1/2}]\\
&&[\sqrt{\frac{m^4 r^7(-4j^2+Mr^3)^2}{[(-2j^3 m^2
+jm^2 r^3(-4m+3r)+\sqrt{m^4(5j^2+Mr^3)(2j^2+r^3(-2M+r))^2}]^2}}].
\label{7}
\end{eqnarray}
Now as standard practice, we can define the Keplerian angular momentum distribution 
$\lambda_K=\frac{L}{E}$. Therefore, corresponding centrifugal force in Hartle Thorne 
geometry can be written as
\begin{eqnarray}
\nonumber
\frac{\lambda_K^2}{r^3}=\frac{m^4 r^4(-4j^2+Mr^3)^2}{[(-2j^3 m^2
+jm^2 r^3(-4m+3r)+\sqrt{m^4(5j^2+Mr^3)(2j^2+r^3(-2M+r))^2}]^2}=F_r.\\
\label{8}
\end{eqnarray}
Thus from above, $F_r$ can be identified as the gravitational force of rotating central object 
having hard surface at the Keplerian orbit in an equatorial plane.
The above expression reduces to Paczy\'nski-Wiita form for $j=0$, $m=1$ and $M=1$. 
Thus, we propose, Eq. (\ref{8}) is the most general form of the gravitational force 
corresponding to the pseudo-potential in accretion disk around rotating central object having 
hard surface in equatorial plane. It needs to be emphasized that though our theoretical result is 
valid for slowly rotating GOHS (HT metric is valid for slowly rotating stars), till date it is 
found observationally that even fast rotating neutron/strange stars has $j$ much less
than $0.5$. Hence, our result will be able to explain the accretion disk properties for 
all observed rotating GOHS. Nevertheless, theoretically $j$ can be much high and for 
any theoretical calculation where $j> 0.5$, our result will not be valid.

\section{Comparison of the Results for Hartle-Thorne geometry and Pseudo-Potential}

Following Mukhopadhyay (2002a), we will check how much correctly the potential 
($V_r=\int F_r dr$) reproduces the values of $r_b$, $r_s$ and $E_s$ for that of
Hartle-Thorne geometry, which can be finally established as a good pseudo-potential in 
accretion disks around any GOHS. As the corresponding equations to calculate $r_b$, 
$r_s$ and $E_s$ are given in Mukhopadhyay (2002a), we are not
repeating those here. Below we enlist corresponding $r_b$, $r_s$ and $E_s$ at 
various values of $j$. 

\vskip0.2cm
{\centerline{\large Table-I}}
{\centerline{\large Values of $r_b$}}
\begin{center}
{
\vbox{
\begin{tabular}{llllllllllllllllllllllllllllllll}
\hline
\hline
$j$ & $0$ & $0.1$ & $0.2$ & $0.3$ & $0.4$ & $0.5$ \\
\hline
\hline
$V_r$  & $4$ & $3.7862$ & $3.5619$   & $3.3255$  &  $3.0762$ & $2.8135$ \\
\hline
HT & $4$ & $3.7954$&  $3.5810$ & $3.3557$  & $3.118$ &  $2.8689$ \\
\hline
\hline
$j$ & $0$ & $-0.1$ & $-0.2$ & $-0.3$ & $-0.4$ & $-0.5$ \\
\hline
\hline
$V_r$  & $4$ & $4.2042$ & $4.4$   & $4.5885$  &  $4.77$ & $4.9462$ \\
\hline
HT & $4$ & $4.1957$&  $4.3838$ & $4.5684$  & $4.7397$ &  $4.9089$ \\
\hline
\hline
\end{tabular}
}}
\end{center}

\vskip0.2cm
{\centerline{\large Table-2}}
{\centerline{\large Values of $r_s$}}
\begin{center}
{
\vbox{
\begin{tabular}{llllllllllllllllllllllllllllllll}
\hline
\hline
$j$ & $0$ & $0.1$ & $0.2$ & $0.3$ & $0.4$ & $0.5$ \\
\hline
\hline
$V_r$ \& HT  & $6.0$ & $5.6648$ & $5.3107$   & $4.9339$  &  $4.5299$ & $4.0934$ \\
\hline
\hline
$j$ & $0$ & $-0.1$ & $-0.2$ & $-0.3$ & $-0.4$ & $-0.5$ \\
\hline
\hline
$V_r$ \& HT  & $6.0$ & $6.3189$ & $6.6240$   & $6.9170$  &  $7.1994$ & $7.4724$ \\
\hline
\hline
\end{tabular}
}}
\end{center}

\begin{table*}[htbp]
{\centerline{\large Table-3}}
{\centerline{\large Values of $E_s$}}
\begin{center}
{
\vbox{
\begin{tabular}{lllllllllllllllllll}
\hline
\hline
$j$ & $0$ & $0.1$ & $0.2$ & $0.3$ & $0.4$ & $0.5$ \\
\hline
\hline
$V_r$  & $-0.0625$ & $-0.0664$ & $-0.0711$   & $-0.0769$  &  $-0.0843$ & $-0.0943$ \\
\hline
HT & $-0.0572$ & $-0.0607$&  $-0.0649$ & $-0.0701$  & $-0.0767$ &  $-0.0857$ \\
\hline
\hline
$j$ & $0$ & $-0.1$ & $-0.2$ & $-0.3$ & $-0.4$ & $-0.5$ \\
\hline
\hline
$V_r$  & $-0.0625$ & $-0.0592$ & $-0.0563$   & $-0.0538$  &  $-0.0516$ & $-0.0497$ \\
\hline
HT & $-0.0572$ & $-0.0542$&  $-0.0517$ & $-0.0494$  & $-0.0474$ &  $-0.0457$ \\
\hline
\hline
\end{tabular}
}}
\end{center}
\end{table*}

From Table-1, it is clear that for all values of $j$, $V_r$ can 
reproduce the value of $r_b$ in very good agreement with general 
relativistic (HT) results. The maximum error in $r_b$ is $\sim 4\%$. Table-2 indicates that 
$V_r$ reproduces $r_s$ with $100\%$ accuracy with Hartle-Thorne geometry. Table-3
shows a maximum possible error in $E_s$ is $\sim 10\%$. Thus,
the potential $V_r$ will produce a slightly larger luminosity than
the general relativistic one in the accretion disk for a particular
value of $j$. It is to be noticed that for counter-rotating central objects, that is for
retrograde solutions, the errors are less than those of co-rotating ones.
Eventually we can tell that our potential can describe more or less all the phenomena quite exactly 
with that of Hartle-Thorne geometry and can claim it as a good pseudo-potential that can
mimic the general relativistic result for the accretion disk around any GOHS, particularly close 
to the equatorial plane.

\section{Conclusions}

In this paper, we have prescribed a pseudo-potential for the modeling of 
accretion disk around rotating central objects with hard surface, which
can be profoundly used to study the fluid dynamics of its disk. For the
non-rotating case, that is for $j=0$, $m=1$ and for $M=1$ the 
potential reduces to well known Paczy\'nski-Wiita potential. The author 
is not apprised of any pseudo-potential derived exclusively for 
rotating hard surfaces, which can mimic its corresponding general relativistic 
metric, as, the studies done using Newtonian potential, Paczynski-Wiita 
potential or other analytical potentials till now are generally meant 
for non-rotating or rotating black holes and not for rotating neutron 
stars.  Following the similar procedure of that of Mukhopadhyay (2002a),
here the potential has been described. As the potential is derived
from the metric itself, its accuracy is much higher. Our potential is valid for both
co-rotating and counter-rotating central objects having hard surface. In fact, the 
counter-rotating results agree better with the general relativistic ones, presumably because
of the larger values of $r_b$ and $r_s$ with respect to that of co-rotating cases. Not only for neutron star,
this potential can be used to study properties of the accretion disks around any 
rotating central objects having hard surface, where general relativistic effect is much profound. Though this  
potential is derived from the corresponding metric itself, yet this is more applicable to study the 
relativistic fluid dynamics of the accretion 
disk around rotating hard surfaces. Other properties of the accretion disk, namely spectral 
studies or epicyclic frequencies may not be analyzed efficiently. If the description of the disk fluid 
properties is acceptable within $10\%$ accuracy, our potential should be well recommended. It must however 
be noticed that as $\theta=\pi/2$ has been chosen at the very beginning of our calculation,
the potential is mainly valid at the equatorial plane. These few hindrances 
notwithstanding, this potential is one of its 1st kind to the best of author's knowledge, in turn describing 
the accretion disk properties around any rotating GOHS in equatorial plane.

Again it has been clearly illucidated in introduction about  
the reasons behind to consider the Hartle-Thorne metric over Kerr metric in the
context of these approximate methods (pseudo-Newtonian potential) to
describe the accretion disk properties around any rotating GOHS in 
equatorial plane. Generalisation of the potential, 
that is a global potential, describing all properties of the accretion disk around rotating
central objects with hard surface may be worked out in future. Next, one should study: how does
the rotation of central objects having hard surface affect the fluid dynamical 
properties in an accretion disk? How does it affect the parameter region of disk? 
In a word, we should study the complete set of global solution of viscous transonic flows around
rotating hard surfaces.




\end{document}